\documentclass[%
 reprint,
 amsmath,amssymb,a
 aps,
 twocolumn
]{revtex4-2} 

\usepackage{graphicx}
\usepackage{dcolumn}
\usepackage{bm}
\usepackage[colorlinks,linkcolor=blue,citecolor=blue]{hyperref}
\usepackage{graphicx}

\usepackage{url}
\usepackage{ulem}
\usepackage{amsmath}
\usepackage{amsfonts}
\usepackage{textcomp}
\usepackage{subfigure}
\usepackage{verbatim}
\usepackage{rotating}

\usepackage{multirow}
\usepackage{booktabs}
\usepackage{lineno}
\usepackage{appendix}

\def \gev  {~\mbox{GeV}}

\def \gevcc{~\mbox{GeV/$c^2$}}

\def \mevcc{~\mbox{MeV/$c^2$}}

\def \jpsi {J/\psi}

\begin{document}

\title{\boldmath Search for sub-GeV dark particles in $\eta\to\pi^0+\rm{invisible}$ decay}

\author{
M.~Ablikim$^{1}$\BESIIIorcid{0000-0002-3935-619X},
M.~N.~Achasov$^{4,c}$\BESIIIorcid{0000-0002-9400-8622},
P.~Adlarson$^{80}$\BESIIIorcid{0000-0001-6280-3851},
X.~C.~Ai$^{85}$\BESIIIorcid{0000-0003-3856-2415},
R.~Aliberti$^{37}$\BESIIIorcid{0000-0003-3500-4012},
A.~Amoroso$^{79A,79C}$\BESIIIorcid{0000-0002-3095-8610},
Q.~An$^{76,62,\dagger}$,
Y.~Bai$^{60}$\BESIIIorcid{0000-0001-6593-5665},
O.~Bakina$^{38}$\BESIIIorcid{0009-0005-0719-7461},
Y.~Ban$^{48,h}$\BESIIIorcid{0000-0002-1912-0374}
% ...
\\
\vspace{0.2cm}
(BESIII Collaboration)\\
\vspace{0.2cm} {\it
$^{1}$ Institute of High Energy Physics, Beijing 100049, People's Republic of China\\
% ...
$^{4}$ Budker Institute of Nuclear Physics SB RAS (BINP), Novosibirsk 630090, Russia\\
% ...
$^{37}$ Johannes Gutenberg University of Mainz, Johann-Joachim-Becher-Weg 45, D-55099 Mainz, Germany\\
$^{38}$ Joint Institute for Nuclear Research, 141980 Dubna, Moscow region, Russia\\
% ...
$^{48}$ Peking University, Beijing 100871, People's Republic of China\\
% ...
$^{60}$ Southeast University, Nanjing 211100, People's Republic of China\\
% ...
$^{62}$ State Key Laboratory of Particle Detection and Electronics, Beijing 100049, Hefei 230026, People's Republic of China\\
% ...
$^{76}$ University of Science and Technology of China, Hefei 230026, People's Republic of China\\
% ...
$^{79}$ University of Turin and INFN, (A)University of Turin, I-10125, Turin, Italy; (B)University of Eastern Piedmont, I-15121, Alessandria, Italy; (C)INFN, I-10125, Turin, Italy\\
$^{80}$ Uppsala University, Box 516, SE-75120 Uppsala, Sweden\\
% ...
$^{85}$ Zhengzhou University, Zhengzhou 450001, People's Republic of China\\

\vspace{0.2cm}
$^{\dagger}$ Deceased\\
% ...
$^{c}$ Also at the Novosibirsk State University, Novosibirsk, 630090, Russia\\
% ...
$^{h}$ Also at State Key Laboratory of Nuclear Physics and Technology, Peking University, Beijing 100871, People's Republic of China\\
% ...
}

}


\begin{abstract}
Using (10087$\pm$44)$\times$10$^{6}$ $J/\psi$ events collected with the BESIII detector at the BEPCII collider at the center-of-mass energy of $\sqrt{s}=3.097~\rm{GeV}$, we report the first search for $\eta\to\pi^0S\to\pi^0\chi\bar{\chi}$ with $S$ denotes an on-shell dark scalar boson and $\chi$ an invisible dark matter particle. No significant signals are observed with $S$ mass ranging from 0 to 400 $\rm{MeV}/c^2$. The upper limits on the branching fractions and the new physics coupling strengths between $S$ and quarks are set to be $(1.8\sim5.5)\times10^{-5}$ and $(1.3\sim3.2)\times10^{-5}$ at the 90\% confidence level, respectively.  The constraints on the dark-matter-nucleon scattering cross section is improved by approximately 5 orders of magnitude over previous dark-matter-nucleon scattering experiments, providing unique insights into sub-GeV dark matter.

\end{abstract}

\oddsidemargin  -0.2cm
\evensidemargin -0.2cm
\newcommand{\BESIIIorcid}[1]{\href{https://orcid.org/#1}{\hspace*{0.1em}\raisebox{-0.45ex}{\includegraphics[width=1em]{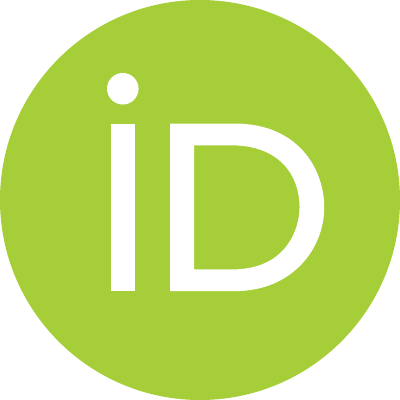}}}}
\maketitle

Dark matter~(DM) constitutes a major component of the Universe,
accounting for 84\% of the total matter content, yet it can not be explained
by the Standard Model~(SM) of particle physics~\cite{Bertone:2016nfn,Cirelli:2024ssz}. While there is indirect evidence supporting the existence of DM~\cite{Planck:2018vyg}, its fundamental nature---including its mass and interactions beyond the gravitational one---remains unclear.
Conventional approaches for the direct detection of DM often rely on nuclear recoils to search for elastic scattering events between DM particles and the target nuclei. Strong constraints have been established for GeV-scale DM~\cite{PandaX:2022xqx,XENON:2024hup,LZ:2022lsv,PandaX:2024qfu,XENON:2025vwd}.
However, sub-GeV DM, which requires the presence of light mediators
below the weak
scale~\cite{Lanfranchi:2020crw,Pospelov:2007mp,Pospelov:2008jd,Arkani-Hamed:2008hhe},
suffers from insufficient recoil energy to exceed the detection
threshold due to its non-relativistic velocities limited by the gravitational potential of the Galaxy~\cite{Drukier:1986tm}. Consequently, traditional direct detection methods are ineffective for sub-GeV DM.

Several innovative approaches have been proposed to enhance the sensitivity of the direct detection to the sub-GeV DM, such as utilizing novel relativistic sub-GeV DM sources and  alternative detection principles.
The new DM sources include cosmic-ray boosted DM (CRDM), where light DM gains kinetic energy through scattering with cosmic rays~\cite{Bringmann:2018cvk,Bondarenko:2019vrb,Ge:2020yuf,Cappiello:2019qsw,PROSPECT:2021awi,PandaX-II:2021kai,CDEX:2022fig,LZ:2025iaw}, and meson-decay boosted DM (MDDM) that is produced from the decays of atmospheric light mesons generated in cosmic-ray interactions~\cite{Alvey:2019zaa,Flambaum:2020xxo,Su:2022wpj,PandaX:2023tfq}.
The new detection principles include utilizing the Migdal effect (MEDM)~\cite{migdal1941ionization,Ibe:2017yqa,XENON:2019zpr,SENSEI:2020dpa,DarkSide:2022dhx,PandaX:2023xgl,SENSEI:2023zdf} and Bremsstrahlung~\cite{Kouvaris:2016afs,XENON:2019zpr}, which transform the signal characteristics from nuclear recoils to electronic recoils, thus achieving a lower detection threshold.
Despite these advances, current experimental sensitivity to sub-GeV DM remains substantially more limited than for GeV-scale DM.

\begin{figure}[htbp] \centering
	\setlength{\abovecaptionskip}{-1pt}
	\setlength{\belowcaptionskip}{10pt}

        \subfigure[]
        {\includegraphics[width=0.225\textwidth]{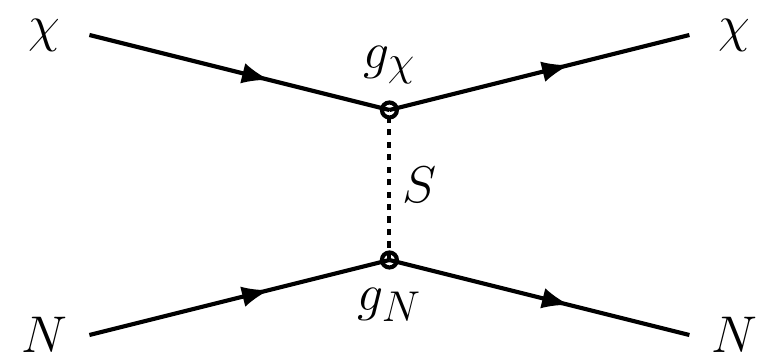}}
        \subfigure[]
        {\includegraphics[width=0.25\textwidth]{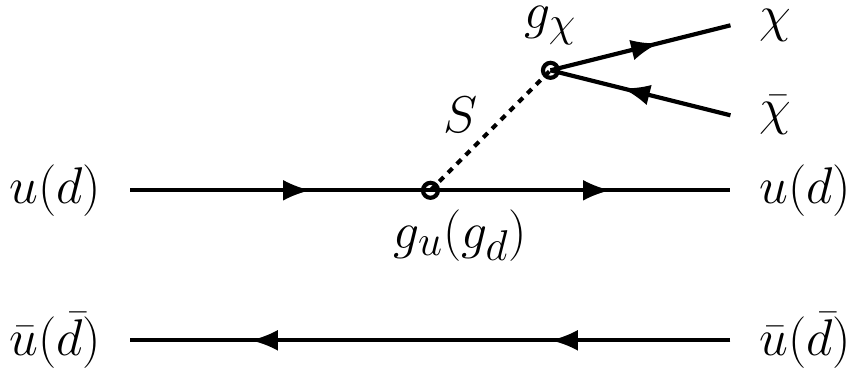}}

	\caption{Feynman diagrams for the DM-nucleon scattering (a) and the $\eta\to\pi^0S\to\pi^0\chi\bar{\chi}$ decay (b). $g_u$, $g_d$, and $g_\chi$ are the coupling strengths, which will be discussed in Eq.~(\ref{eq:operator}). $g_N$ is the effective coupling strength, which is a function of $g_u$ or $g_d$, as will be discussed in Eq.~(\ref{eq:XS}).}
	\label{fig:feynman}
\end{figure}

Collider experiments are also essential for the search for DM, independent of the DM sources and propagation models.
The Large Hadron Collider experiments have established stringent constraints on GeV-scale DM, with sensitivities competitive with those from direct detection experiments~\cite{DeRoeck:2024fjq,ATLAS:2024kpy}.
In this letter, we present a search for sub-GeV DM particles in the meson decay of $\eta\to\pi^0S\to\pi^0\chi\bar{\chi}$ for the first time, where $\chi$ represents the invisible Dirac fermion DM and $S$ denotes the dark scalar boson that mediates the new physics~(NP) interaction between the SM particles and DM.
Typically, new scalar bosons are often assumed to have Higgs-like couplings, resulting in the search for couplings to light states receiving relatively little experimental attention. However, some theoretical studies have found that a flavor-specific scalar boson~\cite{Batell:2018fqo,Delaunay:2025lhl,Batell:2017kty,Batell:2021xsi} can still have a relatively sizable coupling to light quarks, highlighting the potential for NP in light meson decays.
The $\eta$ particle, as an unflavored and narrow-width boson, offers excellent sensitivity to NP decays. The NP interaction in $\eta\to\pi^0S\to\pi^0\chi\bar{\chi}$ is the same as the DM--nucleon interaction, as shown in Fig.~\ref{fig:feynman}, making it comparable to the direct detection experiment. This process is also the most favorable channel of MDDM~\cite{Alvey:2019zaa,Flambaum:2020xxo,Su:2022wpj,PandaX:2023tfq}.

\vspace{-0.0cm}
\vspace{-0.0cm}

The search is based on $(10087\pm44)\times10^{6}$~$J/\psi$ events~\cite{BESIII:2021cxx} collected with the BESIII detector at the BEPCII collider at the center-of-mass energy of $\sqrt{s}=3.097~\rm{GeV}$. The search focuses on the parameter space  $m_S>2m_{\chi}$ with on-shell $S$ in the mass region from $0$ to $400\mevcc$. Considering that the coupling strength between $S$ and $\chi$ is much larger than that between $S$ and the SM quarks, $S$ will predominantly decay into a $\chi\bar{\chi}$ pair, with an invisible branching fraction~(BF) of approximately 100\%.
The $J/\psi\to\phi\eta$ channel is chosen to select the $\eta$ sample due to its low background and high production yield. The signal event candidates are first selected via
$J/\psi\to\phi\eta$, with $\phi\to K^+K^-$ and $\eta$ tagged inclusively via the recoil.  This data sample is designated the $\eta$ inclusive decay tag~(IDT) sample.
Subsequently, candidates for $\eta\to\pi^0S$ are searched for in the selected IDT sample, where $\pi^0$ decays to $\gamma\gamma$ and $S$ is  invisible. The BF of $\eta\to\pi^0S$ is calculated as
\begin{eqnarray}
\mathcal{B}(\eta\to\pi^0S)=\frac{N_{\rm{sig}}}{\mathcal{B}(\pi^0\to\gamma\gamma)\cdot N_{\rm{IDT}}\cdot \hat{\epsilon}},
\label{eq:BF}
\end{eqnarray}
where $N_{\rm{sig}}$ is the signal yield of $\eta\to\pi^0S$ in data, $\hat{\epsilon}$ is its signal efficiency over the IDT selection evaluated on the signal MC simulation sample, $\mathcal{B}(\pi^0\to\gamma\gamma)$ is the BF of $\pi^0\to\gamma\gamma$~\cite{pdg:2024}, and $N_{\rm{IDT}}$ is total number of $\eta$ mesons in the IDT sample.

Details about the BESIII detector design and performance are provided elsewhere~\cite{BESIII:2009fln}. The Monte Carlo~(MC) simulated data samples, as described in Ref.~\cite{BESIII:2023fqz}, are used to determine the detection efficiencies and estimate the background contributions.
The signal MC samples for the two-body process $\eta\to\pi^0S$ are
generated under the assumption that the angular distribution of the
final particles is uniform, with $S$ for different mass hypotheses
from 0 to $400\mevcc$ in steps of $10\mevcc$. In this Letter, the result of $m_S=0\mevcc$ applies to the case of $m_S\ll10\mevcc$.

The selection criteria for the IDT sample are as following.
The charged track selection criteria follow those established in Ref.~\cite{BESIII:2023fqz}.
Particle identification for charged tracks combines the measured information in the multilayer drift chamber, time-of-flight system, and electromagnetic calorimeter~(EMC). The combined likelihoods ($\mathcal{L}(i)$) under different particle $i$ ($i=e,~\mu,~\pi,~\rm{and}~K$) hypotheses are obtained. Kaon candidates are required to satisfy $\mathcal{L}(K)>0$ and $\mathcal{L}(K)>\mathcal{L}(x)$ ($x=e,~\mu,~\pi$).
At least two charged tracks identified as kaons with opposite charges are required.
The invariant mass of selected two kaons is required to lie within the $\phi$ mass region of [1.00,1.04]\,$\gevcc$. To select $\eta$ candidates, the recoiling mass of $K^+K^-$ must fall into the region of $0.45\gevcc<M^{\rm{recoil}}_{KK}<0.65\gevcc$.
To improve the momentum resolution of the reconstructed kaons, a kinematic fit is performed~\cite{Yan:2010zze}, constraining the recoiling mass of $K^+K^-$ to the nominal mass of $\eta$~\cite{pdg:2024}. If there are multiple $K^+K^-$ combinations, the one with the minimal $\chi^2$ value from the kinematic fit is selected as the $K^+K^-$ candidate.

\begin{figure}[htbp] \centering
	\setlength{\abovecaptionskip}{-1pt}
	\setlength{\belowcaptionskip}{10pt}

        {\includegraphics[width=0.49\textwidth]{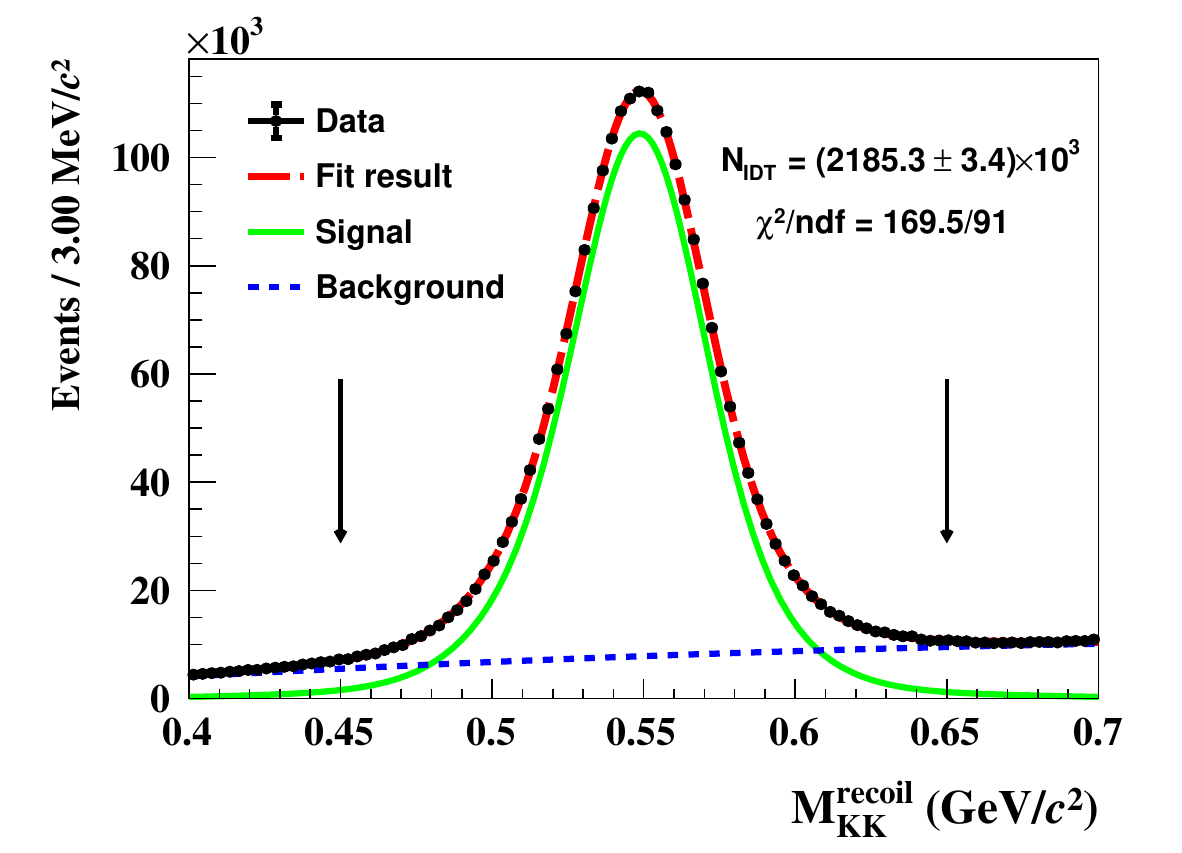}}

	\caption{Fit to the $M^{\rm{recoil}}_{KK}$ distribution. The black points are the data sample, the red dashed line is the fit result, the green solid line is the $\eta$ signal, the blue dashed line is the background. The black arrows indicate the selected region of $M^{\rm{recoil}}_{KK}$.
        }
	\label{fig:fit_IDT}
\end{figure}

The IDT yield is extracted from a binned extended maximum likelihood fit to the distribution of $M^{\rm{recoil}}_{KK}$. In the fit, the signal shape is described by $\mathcal{PDF}^{\rm{IDT}}_{\rm{sig}}\otimes G(\mu^{\rm{IDT}},\sigma^{\rm{IDT}})+f\times G(\mu',\sigma')$,
where $\mathcal{PDF}^{\rm{IDT}}_{\rm{sig}}$ is the signal probability
density function of $M^{\rm{recoil}}_{KK}$ derived from the signal MC
simulation. $G(\mu^{\rm{IDT}},\sigma^{\rm{IDT}})$ is a
Gaussian function with free parameters to describe the resolution
difference between data and MC simulation, and $G(\mu',\sigma')$ is another Gaussian function used to improve the fit quality, with free parameters and coefficient $f$ representing the proportion between two terms of signal shape.
The background shape is described with a second-order polynomial function.
The fit result is shown in Fig.~\ref{fig:fit_IDT}; the fit returns a IDT yield of $N_{\rm{IDT}}=(2185.3\pm3.4)\times10^3$.
This yield includes a small contribution from $\eta$ events arising from $e^+e^-\to K^+K^-\eta$ without the involvement of $\jpsi$ or $\phi$, which are also  utilized in the search for $\eta\to\pi^0S$.

\vspace{-0.0cm}
\vspace{-0.0cm}

The process of interest, $\eta\to\pi^0S$ with $\pi^0\to\gamma\gamma$ and invisible $S$, is further searched for in the IDT sample. The number of the charged tracks is required to be exactly two.
At least two photon candidates are required, with the same identification criteria as in Ref.~\cite{BESIII:2023fqz}.
The invariant mass of the two selected photons, denoted as $M_{\gamma\gamma}$, is required to be in the region of [0.115,0.150]\,$\gevcc$.
To improve the momentum resolution of the reconstructed photons, a kinematic fit is performed~\cite{Yan:2010zze}, constraining the invariant mass of $\gamma\gamma$ to the nominal mass of $\pi^0$~\cite{pdg:2024}. If there are multiple $\gamma\gamma$ combinations, the one with the minimal $\chi^2$ value from the kinematic fit is selected as the $\pi^0$ candidate.
To suppress the background with additional photons or $\pi^0$'s, the total energy of photon candidates other than those from $\pi^0$ ($E^{\rm{tot}}_{\rm{oth.\gamma}}$) is required to be less than 0.1~GeV.
To reduce potential background particles flying to the end cap of the detector that cannot be effectively detected~\cite{Li:2024pox,You:2025nlt}, the cosine of the recoiling angle of $K^+K^-\pi^0$ ($\cos\theta^{\rm{recoil}}_{K^+K^-\pi^0}$) is required to lie within the region of $[-0.7,0.7]$.

The remaining background contributions, such as $\jpsi\to\gamma\eta_c\to\gamma K^+K^-\pi^0$, $\jpsi\to\phi\eta\to K^+K^-\pi^0\pi^0\pi^0$, among others, are suppressed with a requirement on the invariant mass of $K^+K^-\pi^0$, $M_{K^+K^-\pi^0}$. The selection criteria are optimized using the Punzi-optimization method~\cite{Punzi:2003bu} based on the MC samples to maximize the signal-to-noise ratio.
The $S$ signals with different masses are anticipated to exhibit distinct distributions of $M_{K^+K^-\pi^0}$, and the optimized selection criteria are expected to differ accordingly.
For $S$ with masses in the ranges of [0,165)\,MeV/$c^2$, [165,305)\,MeV/$c^2$, and [305,400]\,MeV/$c^2$, the $M_{K^+K^-\pi^0}$ criteria are set to be less than 2.7, 2.15, and $2.0\gevcc$, and the selected samples are called Type I, II, and III, respectively.
With the above selection criteria, the signal efficiency is estimated from the signal MC samples, varying from 17.55\% to 44.28\% for $S$ with different masses.

\vspace{-0.0cm}
\begin{figure*}[htbp] \centering
	\setlength{\abovecaptionskip}{-1pt}
	\setlength{\belowcaptionskip}{10pt}

        \subfigure[]
        {\includegraphics[width=0.329\textwidth]{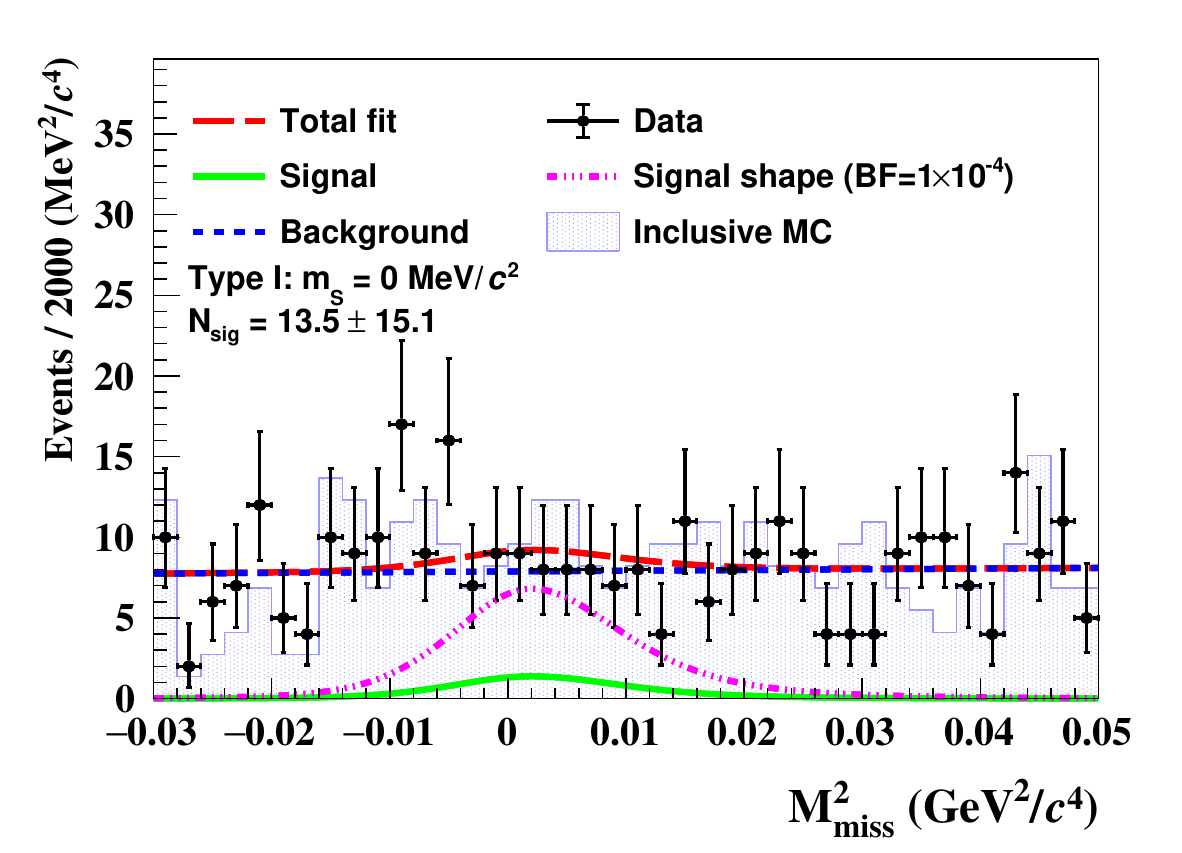}}
        \subfigure[]
        {\includegraphics[width=0.329\textwidth]{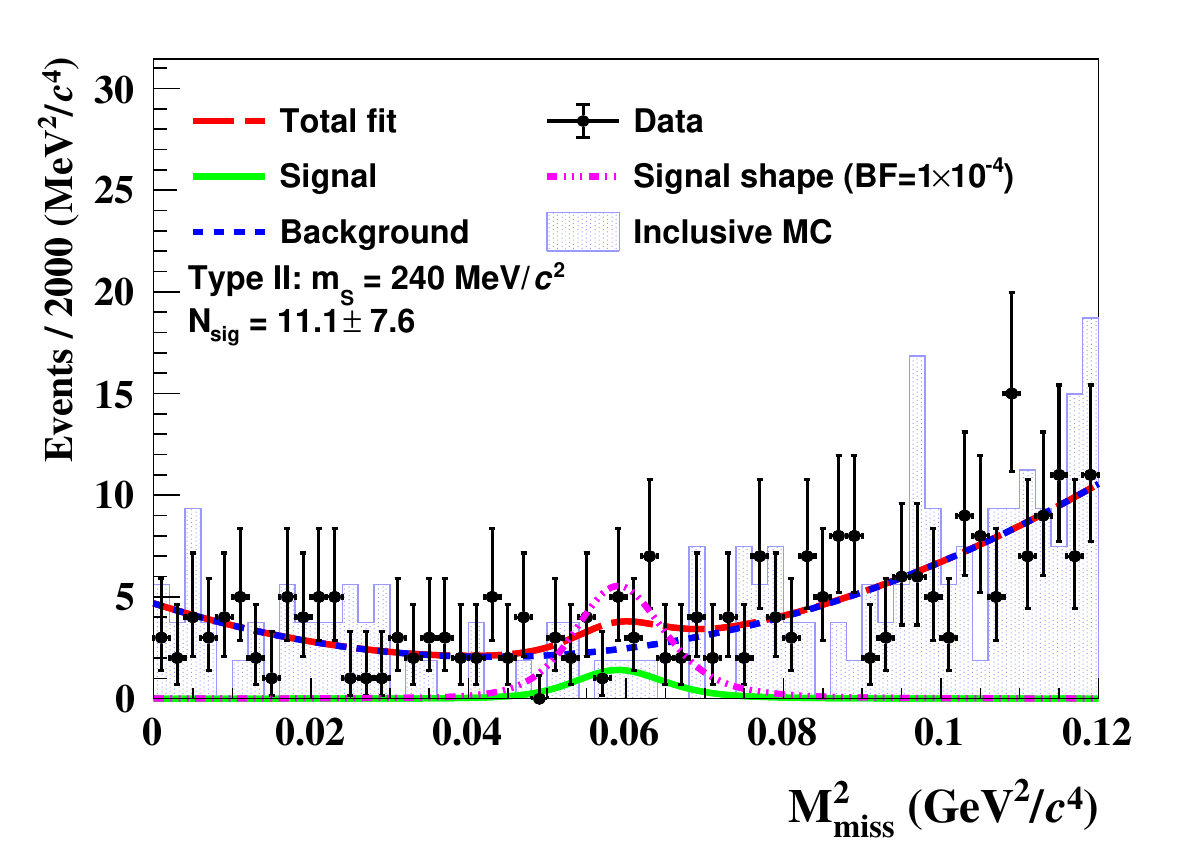}}
        \subfigure[]
        {\includegraphics[width=0.329\textwidth]{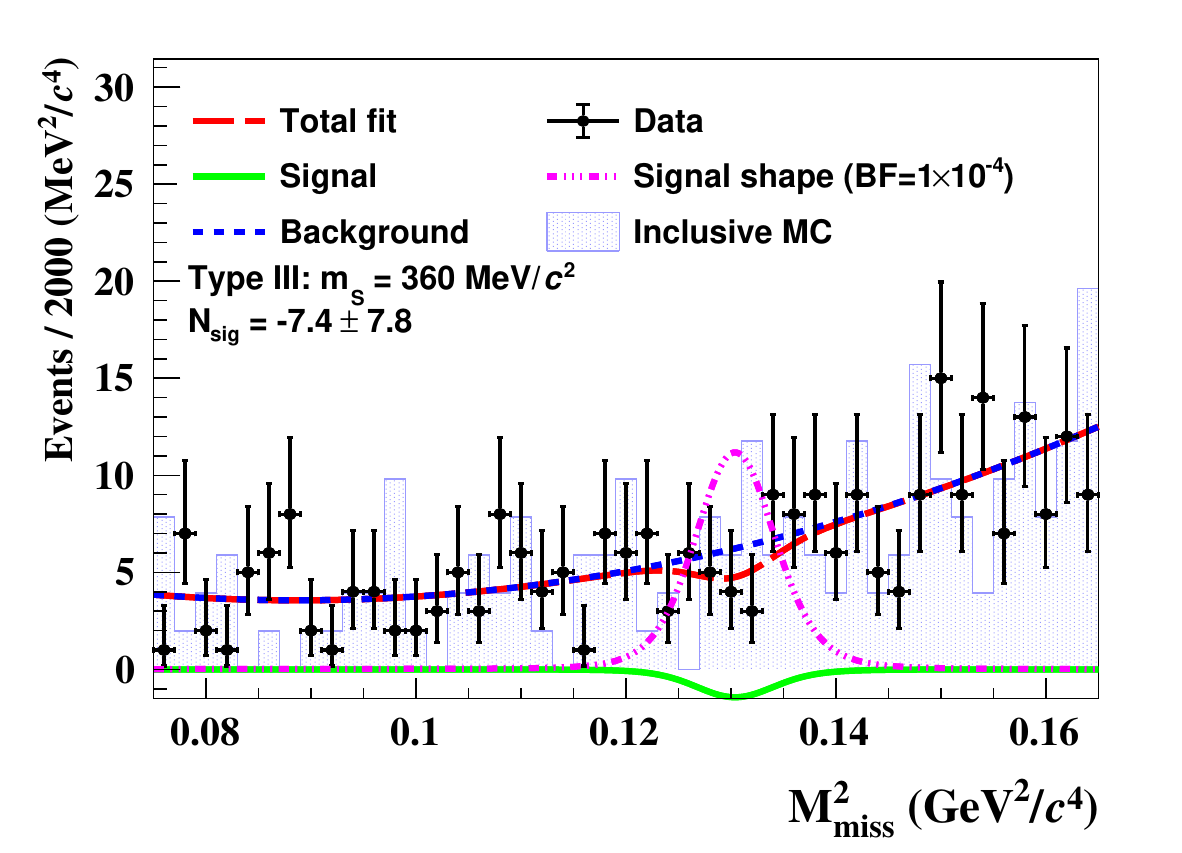}}

	\caption{Fits to the $M^2_{\rm{miss}}$ distributions of the selected samples of Type I (a), Type II (b), and Type III (c). The black points with the error bars are the data sample, the red dashed line is the fit result, the green solid line is the signal, the blue dashed line is the background, the blue shaded histogram is the inclusive MC sample, and the magenta dashed line is the signal shape with the BF of $1\times10^{-4}$. In the fit, $m_S=$ 0 MeV/$c^{2}$ (a), 240 MeV/$c^2$ (b), and 360 MeV/$c^2$ (c) are taken as the examples.}
	\label{fig:fit_DT}
\end{figure*}
\vspace{-0.0cm}

The number of signal events of $\eta\to\pi^0S$ is extracted from an unbinned extended maximum likelihood fit to the distribution of the square of the missing mass, defined as:
\begin{eqnarray}
M^2_{\rm{miss}}=(E_{\rm{c.m.s.}}-\sum_i E_i)^2/\it{c}^{\rm{4}}-(\sum_i \overrightarrow{P}_i)^{\rm{2}}/\it{c}^{\rm{2}},
\label{eq:Mmiss2}
\end{eqnarray}
where $E_{\rm{c.m.s.}}$ is the center-of-mass energy, and $E_i$ ($\overrightarrow{P}_i$) represents the energy (momentum) of particle $i$ from the kinematic fit in the center-of-mass frame, with $i = K^+,K^-,\pi^0$.
The distributions of $M^2_{\rm{miss}}$ for the three types of selected candidates are shown in Fig.~\ref{fig:fit_DT}.
In the fit, the signal shape is described by the simulated shape convolved with a Gaussian resolution function $G(\mu,\sigma)$, where $\mu$ and $\sigma$ are constrained (referred to as the Gaussian constraint) to the values estimated from the control sample of $J/\psi\to\phi\eta\to K^+K^-\gamma\gamma$.
The decay width of $S$, dominated by $S\to\chi\bar{\chi}$, is neglected compared to the experimental resolution.
The background shape is described with a first-order polynomial function for the selected sample of Type I and with a second-order polynomial function for the selected samples of Type II and III.
The signal and the background yields are all free in the fit.

The fit results are presented in Fig.~\ref{fig:fit_DT}, with examples for the masses of $S$ set to 0~MeV/$c^2$ (a), 240~MeV/$c^2$ (b), and 360~MeV/$c^2$ (c). The maximum signal significance ($m_S=240\mevcc$) is found to be less than $2\sigma$, indicating that no significant signal candidates have been observed.
The signal significance is calculated by $\sqrt{2\ln(\mathcal{L}/\mathcal{L}_0)}$, where $\mathcal{L}$ is the maximum likelihood obtained from the fit, and $\mathcal{L}_0$ is the maximum likelihood for $N_{\rm{sig}}=0$.

The systematic uncertainties associated with the IDT yield are discussed below. The uncertainty from the IDT selections can be canceled in this analysis.
The statistical uncertainty of the IDT yield is assigned as 0.2\%.
In the fit to the $M^{\rm{recoil}}_{KK}$ distribution, the systematic uncertainty from the signal shape is estimated by changing the additional Gaussian function to a double Gaussian function in the signal shape.
The systematic uncertainty arising from the background shape is estimated by replacing the shape of the second-order polynomial function with the third-order one.
The systematic uncertainty from the binned fit is assessed by using varying bin sizes. The change in the fitting result for the former case, 0.6\%, is taken as the corresponding systematic uncertainty, while the changes for the latter two cases are negligible.
The total uncertainty of the IDT yield is determined to be 0.7\%.

The systematic uncertainties associated with the signal selection are discussed below.
The uncertainty from the BF of $\pi^0\to\gamma\gamma$ is neglected~\cite{pdg:2024}.
The uncertainty of photon detection is found to be 0.5\% per photon studied using the control sample of $e^+e^-\to\gamma\mu^+\mu^-$. The uncertainty associated with the $M_{\gamma\gamma}$ requirement is assigned as 0.5\% using the control sample of $D^0\to\pi^+\pi^0e^-\nu_e$.
The uncertainty of the $E^{\rm{oth}}_{\rm{oth.\gamma}}$ and $\cos\theta^{\rm{recoil}}_{K^+K^-\pi^0}$ requirement is estimated from the control sample of $\eta\to\gamma\gamma$, which is 1.5\% and 0.1\%, respectively.
The values of these uncertainties are determined by assessing the efficiency difference between the data and the MC simulation in the control samples.
The uncertainty associated with the $M_{K^+K^-\pi^0}$ selection is studied by smearing the $M_{K^+K^-\pi^0}$ spectra of the signal MC samples with a Gaussian function $G(\mu'',\sigma'')$, setting the large parameters $\mu''=\pm5\mevcc$ and $\sigma''=10\mevcc$. The maximum signal efficiency difference, ranging from 0.5\% to 1.4\% for $S$ with different masses, is taken as the conservative uncertainty.
The total systematic uncertainty other than the fitting strategy is calculated by summing up all sources in quadrature, yielding from 2.1\% to 2.5\% for $S$ with different masses.
To consider the uncertainty of the fitting strategy related to the signal and background shapes, we relax the Gaussian constraint on the smear parameters for the signal shape and increase the order of the polynomial function by one to describe the background shape. The most conservative result is assigned as the alternative fit.

\begin{figure}[htbp] \centering
	\setlength{\abovecaptionskip}{-1pt}
	\setlength{\belowcaptionskip}{10pt}

        {\includegraphics[width=0.49\textwidth]{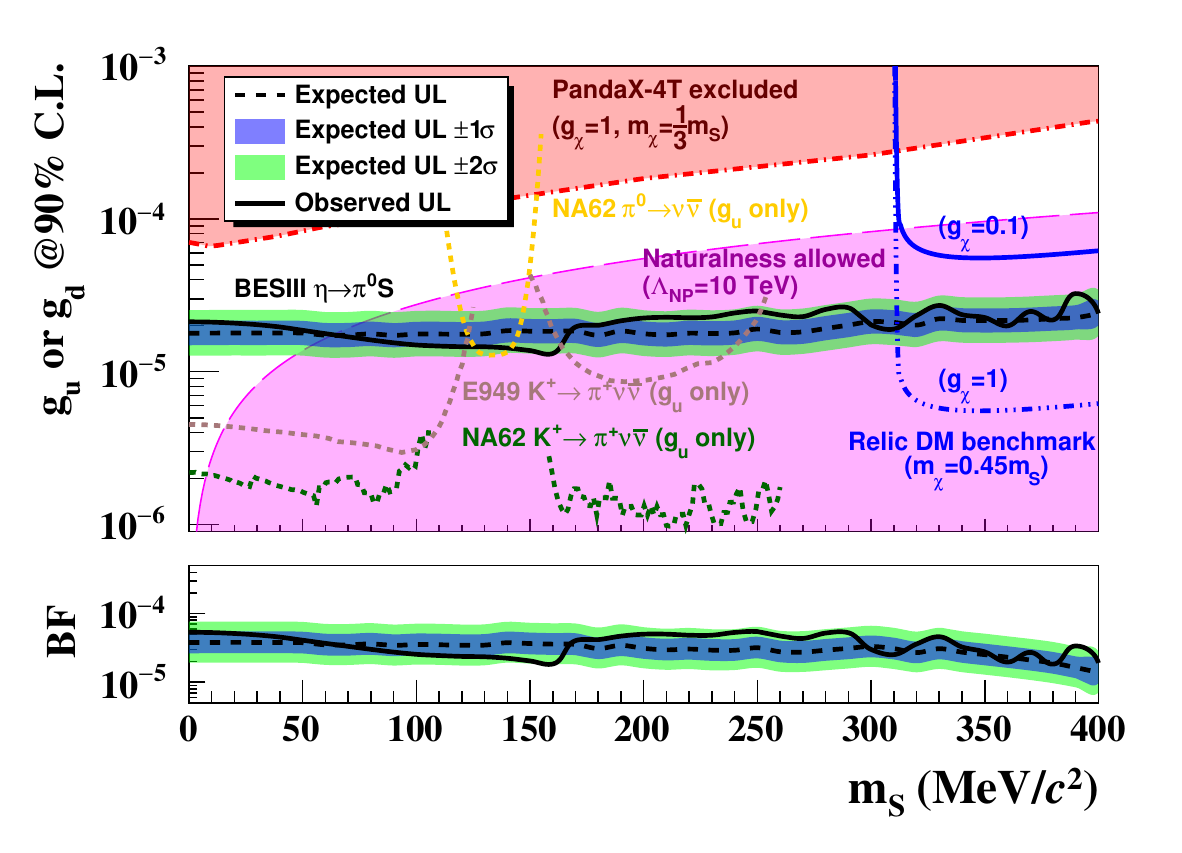}}

	\caption{The ULs on the coupling strengths $g_u$ or $g_d$ (up) and the BFs of $\eta\to\pi^0S$ (bottom) for $S$ with different masses. The black solid (dashed) line is the observed (expected) UL, the blue (green) filled region is the expected UL $\pm1\sigma$ ($\pm2\sigma$) from the MC simulation of zero signal assumption, the red dashed line denotes the published UL from PandaX~\cite{PandaX:2023tfq}, and the purple-filled region represents the naturalness allowed space in the EFT with $\Lambda_{\rm{NP}}=10~\rm{TeV}$.
    The yellow, brown, and dark green dashed lines are the recast ULs from the search for $K^+ \to \pi^+ + \text{invisible}$~\cite{NA62:2025upx,BNL-E949:2009dza,NA62:2020pwi}, which apply only to $g_u$. The blue lines illustrate the thermal relic DM benchmarks for $g_{\chi} = 0.1$ and $g_{\chi} = 1$.
        }
	\label{fig:UL}
\end{figure}

Since no significant excess of signal above the background is observed, upper limits~(ULs) on the BFs are determined using a Bayesian approach following Ref.~\cite{BESIII:2023fqz}, where the BF is calculated by Eq.~(\ref{eq:BF}) and the systematic uncertainty is considered with the method in Refs.~\cite{Liu:2015uha,Stenson:2006gwf}. We set the ULs on $S$ with mass ranging from 0 to 400~MeV/$c^2$ in steps of 10~MeV/$c^2$, which is less than half of the mass resolution.
Finally, the ULs on the BFs at the 90\% confidence level (CL) are calculated by integrating the likelihood distribution with different signal yield assumptions up to the 90\% region; ULs from $1.8\times10^{-5}$ to $5.5\times10^{-5}$ are obtained for different $S$ masses, as shown in Fig.~\ref{fig:UL}.

Next, we use the flavor-specific model~\cite{Batell:2018fqo,Delaunay:2025lhl,Batell:2017kty,Batell:2021xsi} for a further discussion.
In the flavor-specific model, the $S$ couples the Dirac fermion DM $\chi$ and the SM quark $u$ through the following terms in the Lagrangian~\cite{Batell:2018fqo}:
\begin{eqnarray}
\mathcal{L} \supset -g_{\chi} S \bar{\chi}_L \chi_R - g_u S \bar{u}_L u_R + h.c.,
\label{eq:operator}
\end{eqnarray}
where $g_u$ ($g_\chi$) is the effective coupling strength between $S$ and the SM quark (DM), and $L(R)$ marks the chirality.
In the flavor-specific model, the scalar particle can preferentially couple to the first generation particles of the SM, with a coupling strength not necessarily proportional to the Higgs Yukawa couplings.
Here, we consider the case where $S$ couples to the up quark $u$ as an example, and the discussion of $S$ coupling to the down quark, with a coupling strength of $g_d$, is analogous.
The BF of $\eta\to\pi^0S$ can be expressed as~\cite{Batell:2018fqo}:
\begin{eqnarray}
\mathcal{B}(\eta\to\pi^0S)=\frac{ g^2_u c^2_{S\pi^0\eta } B^2}{ 16\pi m_\eta \Gamma_\eta } \lambda^{\frac{1}{2}}(1, \frac{m^2_S}{m^2_{\eta}}, \frac{m^2_{\pi^0}}{m^2_{\eta}}),
\label{eq:BF_the}
\end{eqnarray}
where $c_{S\pi^0\eta} = \frac{1}{\sqrt{3}} \cos\theta - \sqrt{\frac{2}{3}} \sin\theta$, parameterizing the effects from $\eta-\eta'$ mixing with $\theta = -20^{\circ}$, $B=\frac{m^2_{\pi}}{m_u+m_d}=2.6\gevcc$, and $\lambda(a,b,c)=a^2+b^2+c^2-2ab-2ac-2bc$.
The decay rate of $\eta\to\pi^0S$ is sensitive to NP, and the ULs of $\mathcal{B}(\eta\to\pi^0S)$ can provide stringent constraints on $g_u$, as shown in Fig.~\ref{fig:UL}.
Compared to the previously published result from PandaX-4T~\cite{PandaX:2023tfq}, which measured the DM-nucleon scattering mediated by $S$, the constraints from this work are stronger by a factor of 3.3 to 18.3.
In addition, the constraint from the DM-nucleon scattering experiment depends on several other free NP parameters such as $g_{\chi}$ and $m_{\chi}$, while our result of the decay $\eta\to\pi^0S$ can provide a direct constraint on $g_u$.
The search for $ K^+ \to \pi^+ + \text{invisible} $ from $ K^+ \to \pi^+ \nu \bar{\nu} $ analysis in NA62~\cite{NA62:2025upx} and E949~\cite{BNL-E949:2009dza}, and from $ \pi^0 \to \nu \bar{\nu} $ (where the $ \pi^0 $ originates from $ K^+ \to \pi^+ \pi^0 $) analysis in NA62~\cite{NA62:2020pwi}, can provide some constraints on $ g_u $ in the low $ m_S $ region~\cite{Batell:2018fqo}, which is also illustrated for comparison. It is important to note that $ K^+ \to \pi^+ + \text{invisible} $ is sensitive only to $ g_u $~\cite{Batell:2018fqo}, while $ \eta \to \pi^0 S $ is sensitive to both $ g_u $ and $ g_d $.

In Fig.~\ref{fig:UL}, the naturalness relationship is from the effective field theory~(EFT) of $g_u \leq \frac{16\pi^2}{\sqrt{2}} \frac{m_S\nu}{\Lambda^2_{\rm{NP}}}$~\cite{Batell:2017kty},
where $\Lambda_{\rm{NP}}$ is the NP energy scale, $\nu=246.2\gev$ is the Higgs vacuum expectation value, and this relationship is solely used to establish the naturalness bound.
The NP energy scale is set to be 10~TeV, illustrating that the sensitivity of the $\eta\to\pi^0S$ decay at low energy can probe the allowed space from a high NP energy scale. 
Considering the annihilation of $\chi\bar{\chi} \to S \to \pi\pi$ in the cosmic evolution, Fig.~\ref{fig:UL} also shows the thermal relic DM benchmark that accounts for the observed cosmological DM density of $\Omega_{\rm{DM}}h^2 = 0.12$ through the freeze-out mechanism~\cite{Batell:2018fqo,Scherrer:1985zt}, by assuming $m_{\chi} = 0.45 m_S$. For $g_{\chi} = 0.1$, our work can probe and exclude the corresponding thermal relic DM, whereas for $g_{\chi} = 1$, a larger dataset is still needed to further explore the parameter space of thermal relic DM.

\vspace{-0.0cm}
\vspace{-0.0cm}

\begin{figure}[htbp] \centering
	\setlength{\abovecaptionskip}{-1pt}
	\setlength{\belowcaptionskip}{10pt}

        {\includegraphics[width=0.49\textwidth]{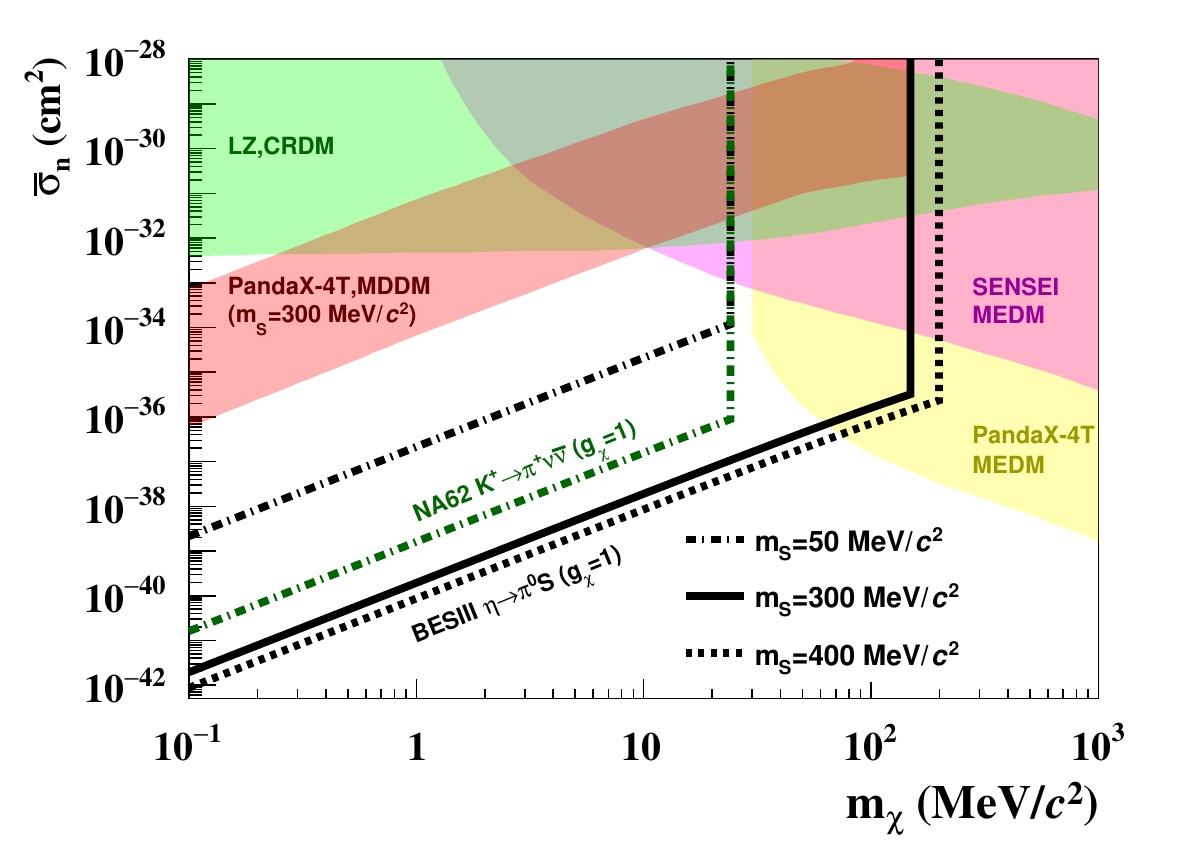}}

	\caption{Constraints on $\bar{\sigma}_n$. The black lines are the constraints from this work with different assumptions, the dark green dashed line is the recast constraint from NA62~\cite{NA62:2025upx}, and the color-filled region is the best excluded region from MDDM~\cite{PandaX:2023tfq}, CRDM~\cite{LZ:2025iaw}, and MEDM~\cite{PandaX:2023xgl,SENSEI:2023zdf} searches. The constraints from BESIII and NA62 apply to the specific model under consideration (Eq.~\ref{eq:operator}).
        }
	\label{fig:XS}
\end{figure}

The sub-GeV DM can interact with the SM quark through a new force mediated by the exchange of $S$~\cite{Flambaum:2020xxo,Alvey:2019zaa,Su:2022wpj}, in addition to the gravitational interaction.
A momentum-independent reference cross-section of the DM-nucleon scattering is expressed as~\cite{PandaX:2023tfq,Flambaum:2020xxo}:
\begin{eqnarray}
\bar{\sigma}_n = \frac{ g_N^2 g^2_{\chi} \mu^2_n }{ \pi (q^2_0 + m^2_S)^2 },
\label{eq:XS}
\end{eqnarray}
where $g_N=(Zy_{Spp} + (A-Z)y_{Snn})/A$, $Z$ and $(A-Z)$ are the numbers of protons and neutrons in the nucleus, respectively;
$y_{Spp} = 0.014 g_u m_p / m_u$, $y_{Snn} = 0.012 g_u m_n / m_u$,
$q^2_0=\alpha^2 m^2_e$ with the fine structure constant $\alpha$, $\mu_n = \frac{ m_{\chi} m_N }{ m_{\chi} + m_N }$, and $m_e,~m_p,~m_n,~m_u$, and $m_N$ represent the masses of the electron, proton, neutron, current up quark, and nucleon, respectively.
By applying the UL of $g_u$ from this work, taking $Z=54$ and $A-Z=77$ for Xenon, and assuming the values of $g_{\chi}$, the constraint on $\bar{\sigma}_n$ is derived, as shown in Fig.~\ref{fig:XS}.
The $g_{\chi}$ is assumed to be 1, which is a sufficiently large value in the EFT, and the constraint can be straightforwardly translated to other values of $g_{\chi}$ using the relation of $\bar{\sigma}_n\propto g^2_{\chi}$. A smaller $g_{\chi}$ would result in a more stringent constraint on $\bar{\sigma}_n$.
Under the similar assumption of $m_S=300\mevcc$ as used by PandaX-4T~\cite{PandaX:2023tfq}, we find that the UL of the DM-nucleon scattering cross-section, resulting from the mediation of $S$, significantly improves the sensitivity of the current direct detection experiment by approximately 5 orders of magnitude.
However, it is essential to note that we do not directly measure the DM-nucleon scattering, and the constraint on the cross-section presented in this work is model-dependent.

\vspace{-0.0cm}
\vspace{-0.0cm}

In summary, we perform the first search for sub-GeV DM in $\eta\to\pi^0S\to\pi^0\chi\bar{\chi}$ with on-shell $S$ using $(10087\pm44)\times10^{6}$~$J/\psi$ events~\cite{BESIII:2021cxx} at the center-of-mass energy of $\sqrt{s}=3.097~\rm{GeV}$.
The ULs on $\mathcal{B}(\eta\to\pi^0S)$ are set to be between $1.8\times10^{-5}$ and $5.5\times10^{-5}$ at the 90\% CL for $S$ with masses from $0$ to $400\mevcc$. Stringent constraints on the coupling strength $g_u$ are derived to be less than $1.3\times10^{-5}$ to $3.2\times10^{-5}$, allowing for exploration of the allowed parameter space from a high NP energy scale and the thermal relic DM.
The constraint on the DM-nucleon scattering cross-section mediated by a dark scalar boson is also derived, with ULs significantly exceeding the sensitivity of the previously published results from the DM-nucleon scattering experiment.
This $\eta\to\pi^0+\rm{invisible}$ search highlights the substantial potential of the BESIII experiment in the study of sub-GeV DM, benefiting from the low background environment and good reconstruction. The future $\eta$ factories, such as REDTOP~\cite{REDTOP:2022slw} and HIAF~\cite{Chen:2024wad}, are expected to collect trillions of $\eta$ candidates but it may be challenging to study such invisible decay at these facilities due to the huge backgrounds.
Our result on $\eta\to\pi^0+\rm{invisible}$ is a great prospect for  future super tau-charm facility~\cite{Achasov:2023gey} planning to collect $10^{12}$ $\jpsi$ events.
Constrained by the phase space, the process of $\eta\to\pi^0S$ can only be used to study $S$ with masses up to approximately 400~MeV/$c^2$, while $\eta'\to\pi^0S$, although with significantly lower sensitivity due to the much larger width of $\eta'$, can extend the mass range up to approximately 800~MeV/$c^2$ in the future analyses using a similar strategy as BESIII.

\textbf{Acknowledgement}

The BESIII Collaboration thanks the staff of BEPCII (https://cstr.cn/31109.02.BEPC) and the IHEP computing center for their strong support, thanks to the helpful discussion with Brian Batell. This work is supported in part by National Key R\&D Program of China under Contracts Nos. 2023YFA1606000, 2023YFA1606704; National Natural Science Foundation of China (NSFC) under Contracts Nos. 125B2107, 11635010, 11935015, 11935016, 11935018, 12025502, 12035009, 12035013, 12061131003, 12192260, 12192261, 12192262, 12192263, 12192264, 12192265, 12221005, 12225509, 12235017, 12361141819; the Chinese Academy of Sciences (CAS) Large-Scale Scientific Facility Program; the Strategic Priority Research Program of Chinese Academy of Sciences under Contract No. XDA0480600; CAS under Contract No. YSBR-101; 100 Talents Program of CAS; The Institute of Nuclear and Particle Physics (INPAC) and Shanghai Key Laboratory for Particle Physics and Cosmology; ERC under Contract No. 758462; German Research Foundation DFG under Contract No. FOR5327; Istituto Nazionale di Fisica Nucleare, Italy; Knut and Alice Wallenberg Foundation under Contracts Nos. 2021.0174, 2021.0299; Ministry of Development of Turkey under Contract No. DPT2006K-120470; National Research Foundation of Korea under Contract No. NRF-2022R1A2C1092335; National Science and Technology fund of Mongolia; Polish National Science Centre under Contract No. 2024/53/B/ST2/00975; STFC (United Kingdom); Swedish Research Council under Contract No. 2019.04595; U. S. Department of Energy under Contract No. DE-FG02-05ER41374.

\bibliographystyle{apsrev4-1}
\bibliography{mybib.bib}


\onecolumngrid

\clearpage
\newpage

\textbf{\boldmath\large Supplemental Material for ``Search for sub-GeV dark particles in $\eta\to\pi^0+\rm{invisible}$ decay"}

\begin{appendices}

\section{Signal yield and signal significance}
\begin{figure*}[htbp] \centering
	\setlength{\abovecaptionskip}{-1pt}
	\setlength{\belowcaptionskip}{10pt}

        \subfigure[]
        {\includegraphics[width=0.49\textwidth]{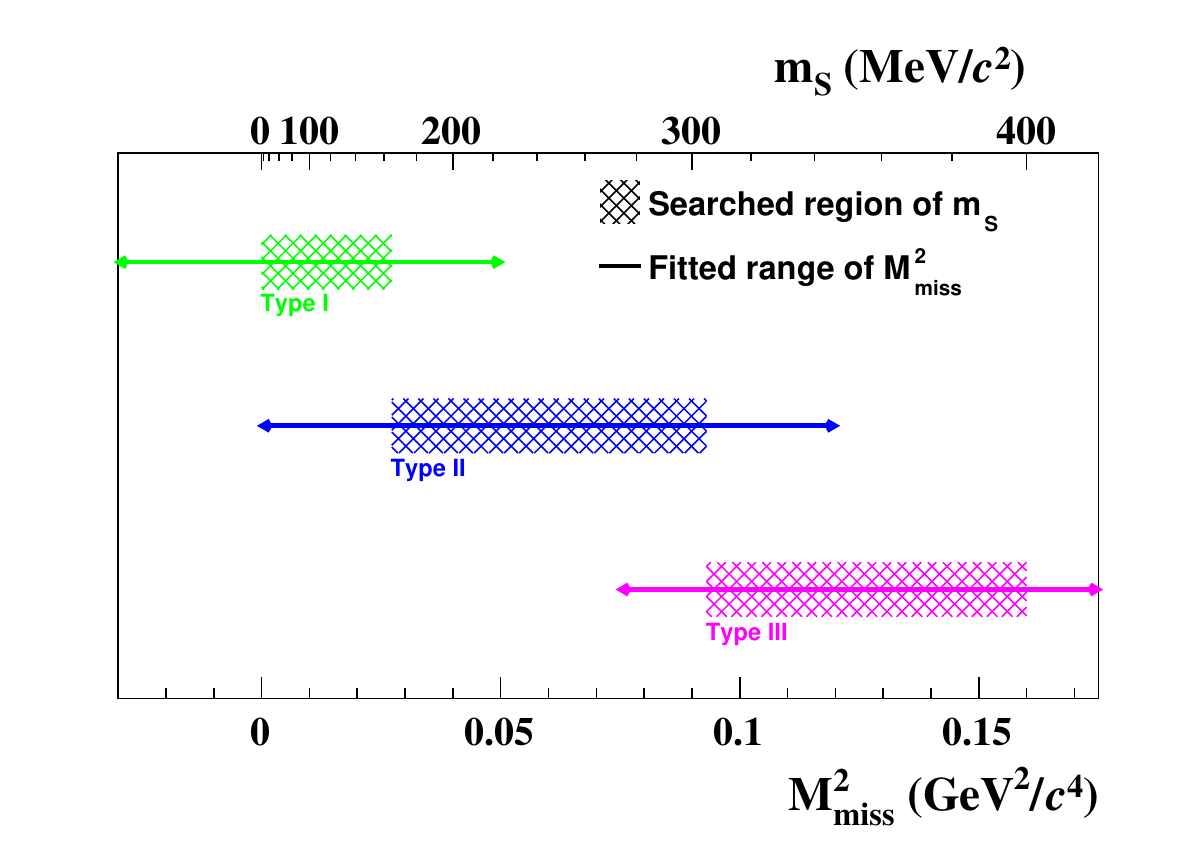}}
        \subfigure[]
        {\includegraphics[width=0.49\textwidth]{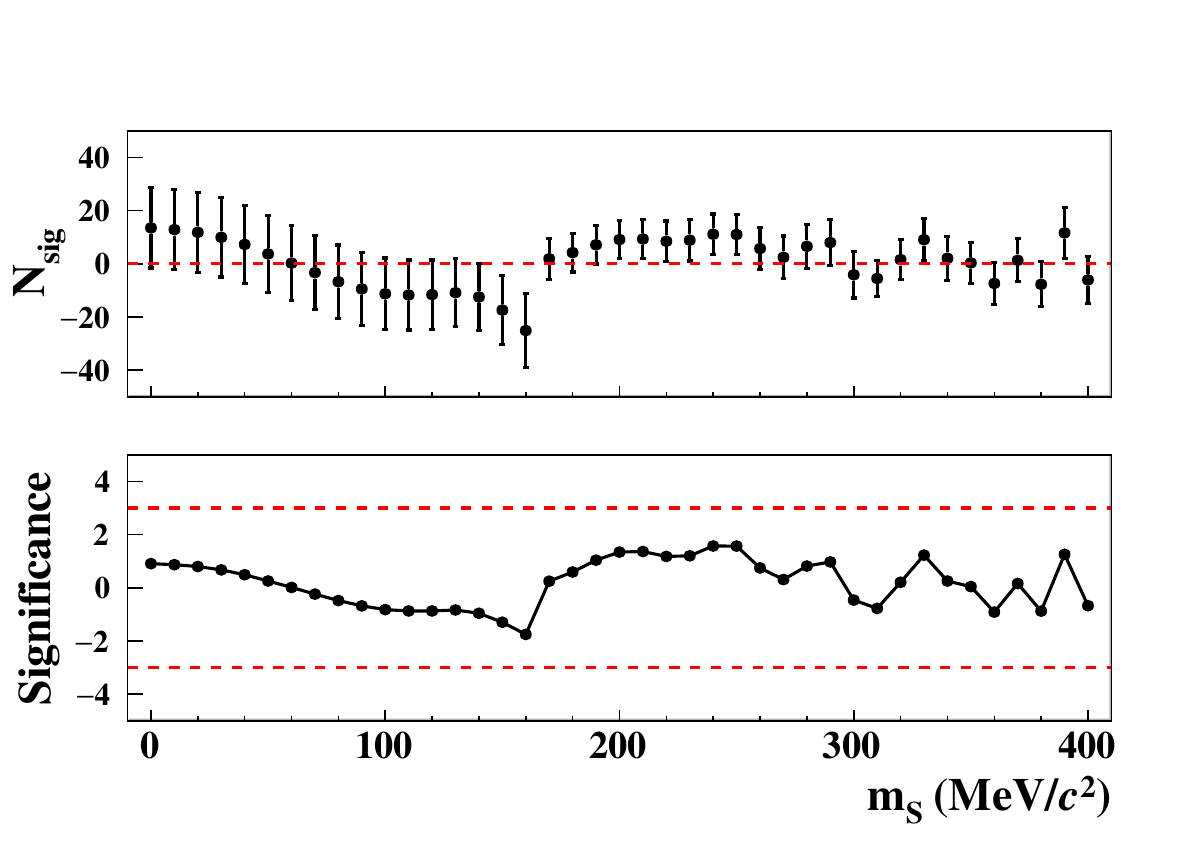}}

	\caption{(a) Fit regions of $M^2_{\rm{miss}}$ corresponding to the searched mass region of $S$ for three types of samples. The shadow regions represent the searched region of $m_S$, while the solid lines with arrows mark the fitted region of $M^2_{\rm{miss}}$. (b) Fitted signal yields $N_{\rm{sig}}$ and the signal significance of $S$ for different mass hypotheses.}
	\label{fig:app_sig}
\end{figure*}

The signal yields of $\eta\to\pi^0S$ are extracted from the distribution of $M^2_{\rm{miss}}$ of the selected samples comprising three different types. The fit regions corresponding to the searched mass region of $S$ for the three types of samples are shown in Fig.~\ref{fig:app_sig} (a). The fitted signal yield and the signal significance are shown in Fig.~\ref{fig:app_sig} (b).

\vspace{-0.0cm}
\vspace{-0.0cm}

\section{Signal efficiency and systematic uncertainty}
\begin{figure}[htbp] \centering
	\setlength{\abovecaptionskip}{-1pt}
	\setlength{\belowcaptionskip}{10pt}

        {\includegraphics[width=0.49\textwidth]{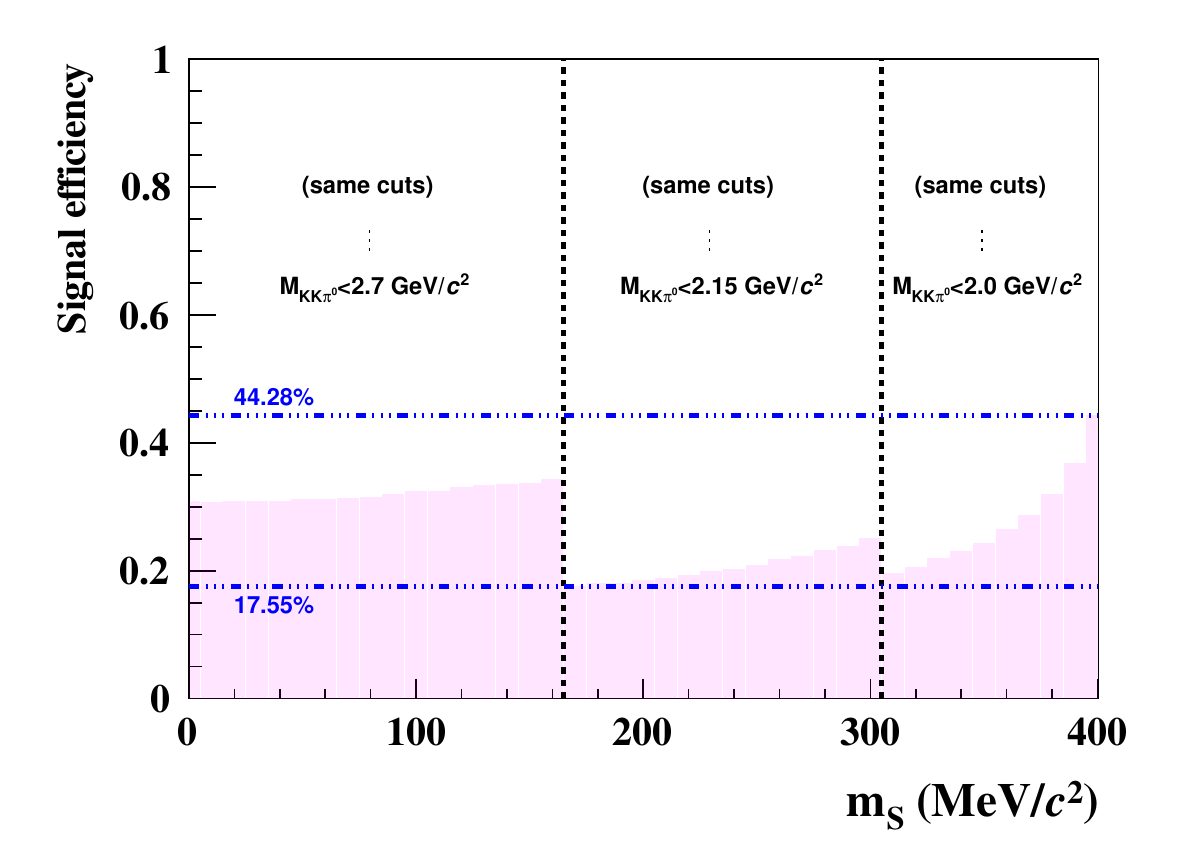}}

	\caption{Signal efficiencies of $S$ for different mass hypotheses. The black dashed lines mark the samples with different selected criteria, and the blue dashed lines mark the maximum and minimum efficiency among $S$ for different mass hypotheses.
        }
	\label{fig:app_eff}
\end{figure}

The signal efficiencies of $S$ for different mass hypotheses are shown in Fig.~\ref{fig:app_eff}. The systematic uncertainties of the signal efficiency associated with the uncertainty of the IDT yield are shown in Table~\ref{tab:sys}.
\vspace{-0.0cm}
\vspace{-0.0cm}

\begin{table}[h]
	\centering
	\caption{Summary of the relative systematic uncertainties in percent for the IDT yield and the signal efficiency. The total value is calculated by summing up all the sources in quadrature.}
        \begin{tabular}{{l}{r}}
		\midrule \midrule
		Source & Uncertainty (\%)\\
		\midrule
		IDT tag yield & 0.7 \\
            Photon detection & 1.0 \\
            $M_{\gamma\gamma}$ requirement & 0.5 \\
            $E^{\rm{tot}}_{\rm{oth}.\gamma}$ requirement & 1.5 \\
            $\cos\theta^{\rm{recoil}}_{K^+K^-\pi^0}$ requirement & 0.1 \\
            $M_{K^+K^-\pi^0}$ requirement & 0.5$\sim$1.4 \\
            \midrule
            Total & 2.1$\sim$2.5 \\
		\midrule \midrule
	\end{tabular}
	\label{tab:sys}
\end{table}

\end{appendices}

\end{document}